# HIGH ENERGY TWO-PHOTON INTERACTIONS AT THE LHC

KRZYSZTOF PIOTRZKOWSKI

*Université catholique de Louvain, Chemin du Cyclotron 2, B-1348 Louvain-la-Neuve*

*E-mail: k.piotrzkowski@fynu.ucl.ac.be*

Two-photon events at the LHC are characterized by the protons scattered at very small angles and the particles centrally produced via the γγ fusion. To select these events from the huge samples of generic *pp* interactions a detection of the scattered protons, or tagging two-photon interactions is necessary. It requires installation of the high-resolution position-sensitive detectors close to the proton beam and far from the interaction point. Efficient measurement of the forward-scattered protons will open a new field of studying high-energy photon-photon interactions at remarkable luminosity, reaching 1% of that in *pp* collisions. In this paper a few aspects of tagging two-photon interactions as well as several most exciting topics in the high-energy two-photon physics at the LHC are presented.

# Introduction

Recently, a method of tagging two-photon interactions at the LHC has been proposed [1]. These interactions, to a good approximation, proceed in two steps: first, the photons are emitted by incoming protons, and then the photons collide producing a system *X*. The method is based on the measurement of the forward scattered protons using detectors similar to those planned for measurements of the elastic *pp* scattering at the LHC [2]. The final state *X* will be detected in the central detectors as in the CMS experiment. Tagging is essential for extraction of high-energy two-photon interactions, at the γγ center of mass energy *W* even beyond 1 TeV. A significant luminosity of the tagged two-photon interactions will allow for new, complementary physics studies at the LHC, in particular for searches for new phenomena. In general two-photon events are cleaner than the *pp* ones and usually the final state is fully contained within acceptance of the central detectors. Particularly interesting are those events where only two or one heavy particle is exclusively produced via the γγ fusion as in the $\gamma\gamma \rightarrow H^0$ case, for example. The event reconstruction can usually be more precise than for the generic *pp* collisions, and it corresponds more to the experimental conditions at the $e^+e^-$ colliders. In fact, also the considered physics topics in high-energy γγ collisions are closely related to the subjects studied at the $e^+e^-$ colliders – in particular to the physics studies planned for the future γγ collider [3]. One should note that of course the photon collider offers much higher luminosity, but a lower energy reach than that available in the two-photon collisions at the LHC. Last but not least, the two-photon physics will be done parasitically to the mainstream physics program, and will complement it, at a very limited incremental cost.

The content of this paper has two major ingredients. First, after a short introduction of the tagging technique several experimental issues are discussed, in particular those relevant for estimates of the effective γγ luminosity. Secondly, the most interesting physics topics are discussed, with emphasis on the exclusive production of the Higgs boson and pairs of the charged SUSY particles.

**Tagging Aspects**

In two-photon processes protons are scattered at angles comparable to the beam angular divergence at the interaction point (IP). The scattered protons can however be measured when a fraction of the initial proton energy, *x*, carried away by a photon is significant. In such a case these protons are more strongly deflected by the beam-line magnets and can be detected in the so-called Roman pots installed far away from the IP and close to the proton beam (see Fig. 1). The detectors are capable of measuring the distance of the scattered proton and its momentum direction with respect to the proton beam at a given location.

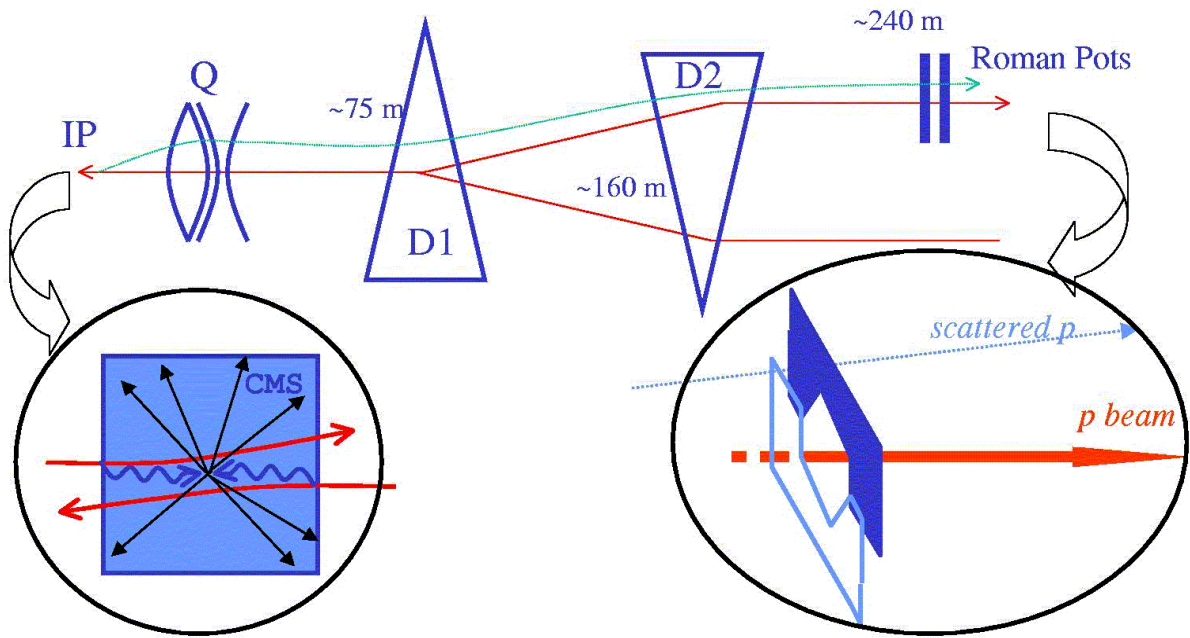

**Figure 1**. Sketch of the LHC beam-line with the proposed localization of the Roman pot detectors for tagging two-photon events; *Q* and *D* symbols stand for the machine quadrupoles and dipoles.

At some 240 m from the IP, the relations between the detector variables and the scattered proton energy and angle at the IP, are particularly simple. The distance in the horizontal plane $\Delta_x$ measures then the proton energy loss, hence the tagged photon energy. Other measured variables, direction of the proton momentum in the horizontal plane $\theta_x$ and distance from the beam axis in the vertical plane $\Delta_y$ are used to reconstruct the proton scattering angle at the IP (from its two projections, $\theta_x^*$ and $\theta_y^*$) as schematically presented is in Tab. 1. The beam dispersion at this location, *D*, is of about 100 mm.

| Variable at the IP | Detector variable @ 240 m |
|---|---|
| $\theta_x^*$ | $3\theta_x$ |
| $\theta_y^*$ | 100 [μrad/mm] $\Delta_y$ |
| $x$ | $\Delta_x/D$ |

**Table 1**. Relations between 'true' variables at the IP and the detector variables according to the LHC beam optics v6.0, where $D \approx 100$ mm.

The photon virtuality $Q^2$ can then be calculated using $Q^2 \cong (1-x)E^2[(\theta_x^*)^2 + (\theta_y^*)^2]$, where *E* is the beam energy, and the photon energy is equal to *xE*.

The tagging efficiency is determined by minimum distance between the detector sensitive edge and the proton beam. For small beam widths, a 1 mm minimum detector approach is usually required to ensure enough space for the beam steering [2]. At the recently advocated detector location at about 240 m [1,4] the horizontal beam width is small and the 1 mm distance corresponds to a minimum tagged photon energy of 70 GeV, that is to 1% of the beam energy. If the maximum tagged energy of 700 GeV and, the maximum virtuality, $Q^2_{max} = 2$ GeV$^2$, of the colliding photons are assumed, one obtains the tagged effective luminosity spectra as a function of *W* [1], see Fig. 2. The *double tagging* corresponds to the case when the two scattered protons are detected, whereas the *single tagging* occurs when only one proton is detected. In such a case, however also those two-photon events are tagged where one proton *dissociates* and does not survive the interaction. In fact, these *inelastic* two-photon events have even higher effective luminosity than the nominal, *elastic* events, and the total luminosity available for the tagged two-photon collisions is significant, reaching 1% of the *pp* luminosity for *W* > 100 GeV. One should also note that the luminosity spectrum extends to a very large *W*, even beyond 1 TeV. The detector approach of 1 mm assumed above in fact requires an edgeless detector, i.e. a detector sensitive right from its mechanical edge. This might be considered too optimistic and instead one can conservatively consider a 2 mm distance between the detector sensitive edge and the beam.

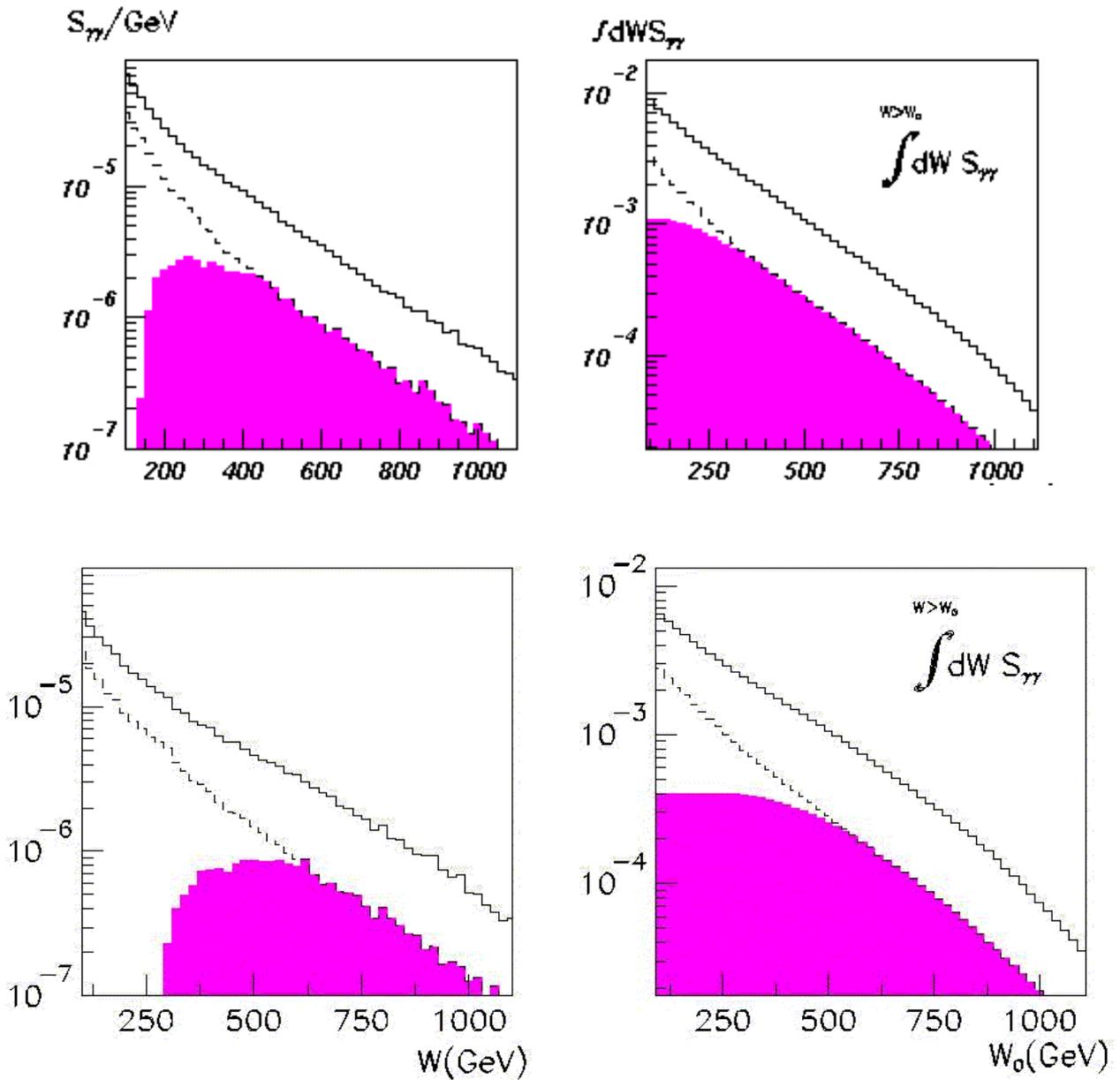

**Figure 2**. Tagged photon-photon luminosity spectrum $S\gamma\gamma$ and its integral $\int_{W_0} dW\, S\gamma\gamma$, assuming double tags (shaded histograms) and single tags, for all events (solid line) and for elastic events (dashed line); above for a 1 mm and below for a 2 mm approach.

In Fig. 2 the luminosity spectra and their integrals (equal to the probability of $\gamma\gamma$ collisions in a *pp* single collision) are shown also for that case, demonstrating that the single-tagged spectrum is modestly affected, whereas the low *W* part of the double-tagged spectrum is as expected suppressed. This shows that the tagging efficiency does not critically depend on the closest approach to the proton beam.

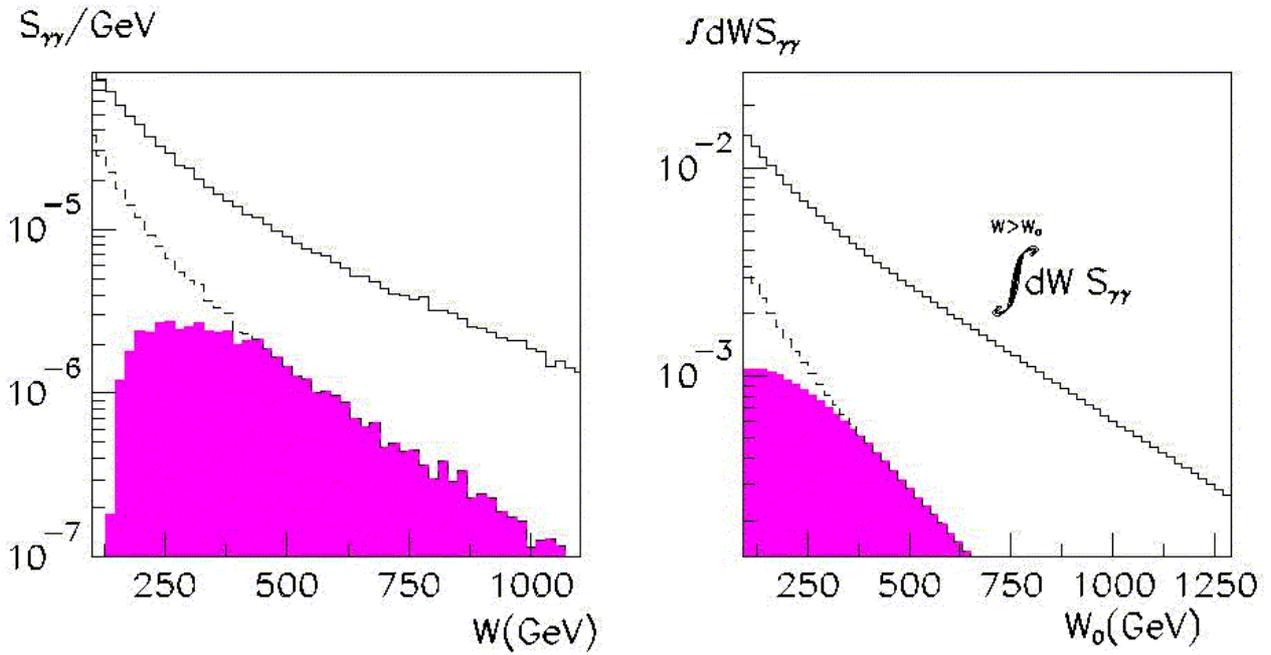

**Figure 3.** Tagged photon-photon luminosity spectrum $S\gamma\gamma$ and its integral $\int_{W_0} dW\, S\gamma\gamma$, assuming double tags (shaded histograms) and single tags, for all events (solid line) and for elastic events (dashed line); all for a 1 mm approach but assuming for the inelastic events $M_N < 7000$ GeV.

So far, following Ref. [1], for the inelastic production the maximum dissociation mass $M_N$ of 20 GeV has been assumed. Allowing for the interactions more inelastic and $M_N$, for example, as large as 7000 GeV significantly increases the effective $\gamma\gamma$ luminosity, see Fig. 3. Virtuality of the photon coupled to the inelastic vertex also increases – for the additional events (with $M_N > 20$ GeV) the average $Q^2$ is about 150 GeV$^2$, resulting in a significant $p_T$ of the system X. Reconstruction of these events is more difficult and requires detailed studies. One should also note that for the high-$Q^2$ events the contribution from the Z exchange is not negligible.

Two-photon interactions in ion collisions are enhanced owing to the coherence effects [5]. At low W the enhancement scales as $Z^4$ but with increasing W it becomes weaker resulting in incoherent production ($\sim Z^2$) at large W. Tagging $\gamma\gamma$ interactions is however not very practical in this case. It is then restricted to a very large $W$ domain where the coherent interactions are much suppressed and γγ luminosity is small.

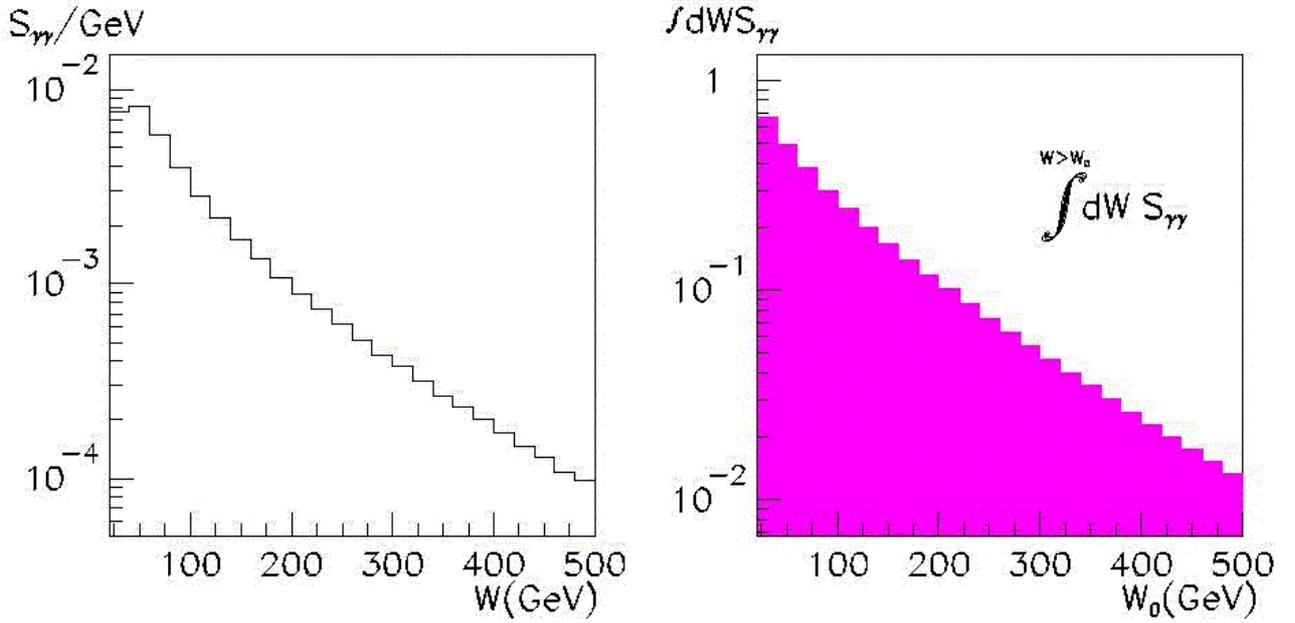

**Figure 4.** Tagged luminosity spectrum $S_{γγ}$ and its integral $\int_{W_0} dW\, S_{γγ}$ for *pAr* collisions and single tagging, assuming elastic production and a 1 mm approach.

Situation is different for proton-ion collisions, where the scattered proton can be detected in the same manner as in the *pp* case, and the single-tagged luminosity is high, especially below $W ≈ 100$ GeV where the enhancement is still significant. For example, almost 50% of the proton-Argon luminosity $L_{pAr}$ is available for γγ collisions at $W > 50$ GeV (see Fig. 4), and $L_{pAr}$ itself might reach values close to $10^{32}$ cm$^{-2}$s$^{-1}$. The luminosity of the tagged γγ events at medium $W$ might be therefore comparable to that in the *pp* collisions. In addition, tagging in *pA* collisions can be used to check γγ event selection in the *AA* collisions. The untagged γγ luminosity in the *ArAr* collisions is high and competitive for $W < 100$ GeV [5].

Finally, two-photon exclusive production of lepton pairs will be an excellent monitoring tool for the tagging efficiency and its energy scale. These events can be selected using a standard CMS di-muon trigger and be used off-line for a number of systematic studies, including the luminosity normalization and contribution of the inelastic production, or the accidental tagging.

**Physics Highlights**

Exclusive production of heavy particles will possibly be the most exciting subject in two-photon research at the LHC. In particular, the Higgs boson might be observed in γγ collisions at the LHC − as can be seen in Fig. 5, it will be statistically limited and will not be a discovery channel, it will however bring new, relevant information. The events are very central, usually within the acceptance of the CMS tracking ($|\eta|<2.5$), therefore high overall detection efficiency is expected. The 'irreducible' background due to the *b*-quark, *W* and *Z* pair production, for the respective decay channels, is comparable to the signal for a few GeV window in the γγ center-of-mass energy.

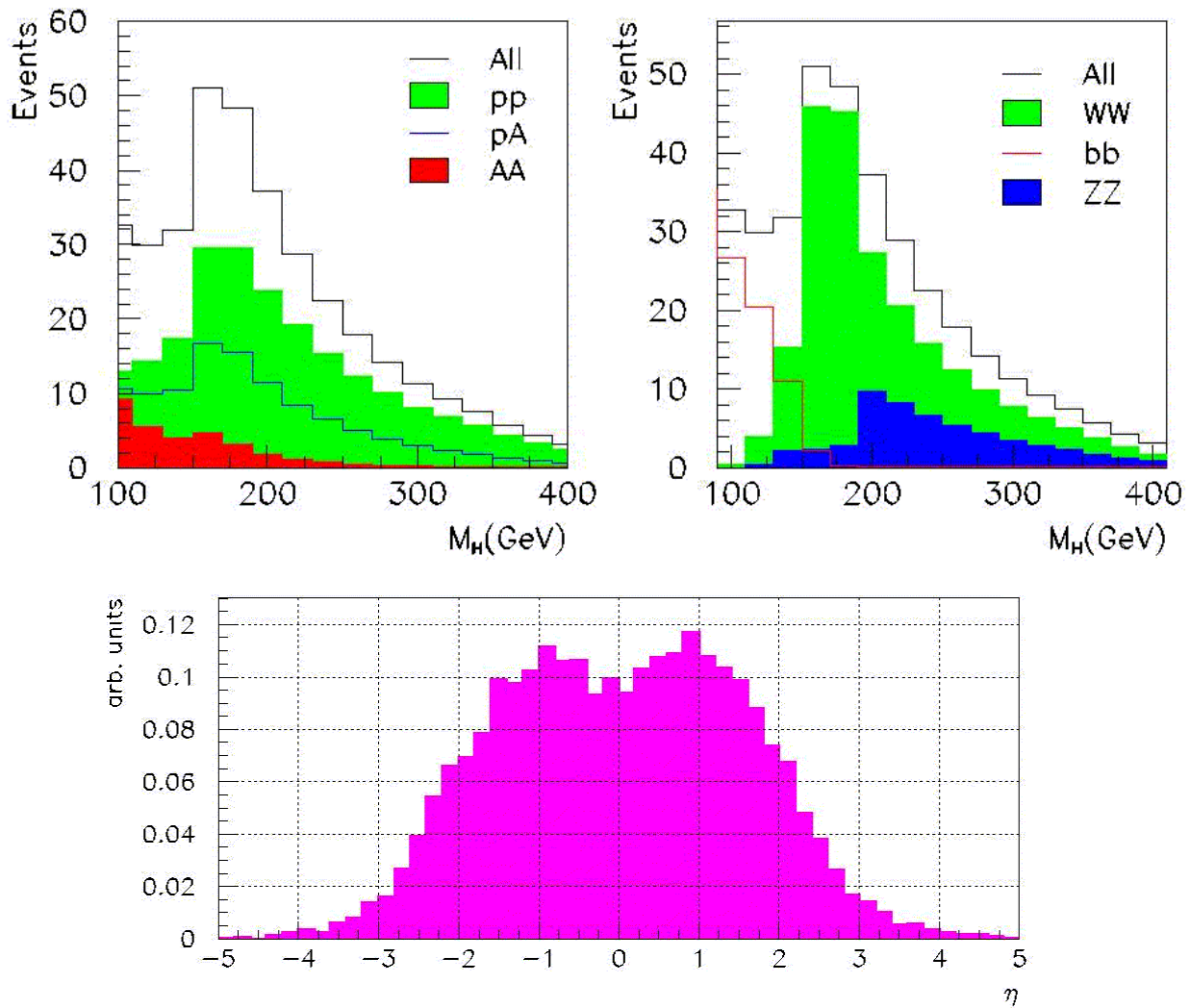

**Figure 5**. (Above) Number of the Standard Model Higgs boson events as a function of its mass, exclusively produced in γγ collisions for the integrated *pp, pA and AA* luminosity of 30 fb$^{-1}$, 300 and 30 pb$^{-1}$, respectively; the event distributions for the major decay modes, $H^0 \rightarrow \bar{b}b$, $W^+W^-$ and $ZZ$ – all branching ratios and widths obtained with the HDECAY 2.0 program [6]. (Below) The pseudo-rapidity distribution of the *b*-quark from the Higgs boson decays for $M_H$ = 120 GeV.

The *W* and *Z* boson pair production in the γγ collisions at the LHC is interesting on its own right. The number of the produced *W* pairs will be similar to that at LEP II but at much higher center-of-mass energy – for example, about 3000 *W*-pairs will be tagged at W > 500 GeV for the integrated *pp* luminosity of 30 fb$^{-1}$. This will allow for precision tests of the γ*WW* coupling and for new physics searches, as for example signatures of Large Extra Dimensions [7]. The *Z* pair two-photon production is much suppressed in the Standard Model but an interesting search for the anomalous γ*ZZ* coupling can also be performed.

The exclusive $t\bar{t}$ two-photon production will be statistically limited – only about one hundred events will be tagged for the 'canonical' luminosity. The observed cross-section, which is proportional to the fourth power of the particle charge, will however directly and precisely measure the top charge.

It has been recently argued [8] that a measurement of the two-photon production of photon pairs at very high transverse momenta at the LHC can be used for searching for the Dirac monopoles. Tagging two-photon events will certainly improve the reconstruction of these events, decrease systematic uncertainties and will help to suppress the backgrounds.

If supersymmetric particles are found at the LHC then the *sparticle* pairs produced in γγ collisions can be used to test the structure and parameters of the underlying theory. The two-photon production of pairs of the charged sparticles is particularly simple – it is a pure QED process. In Fig. 6, the number of the tagged spairs is plotted as a function of the sparticle mass, assuming a simplified scenario of the degenerate sleptons, i.e. the equal masses for the left- and right-handed sleptons [9]. It shows that the charginos might be detected up to the masses of about 300 GeV, and the sleptons and charged Higgs bosons up to the masses of about 200 GeV, covering a significant fraction of benchmark spectra of the minimal SUSY [10]. Studies of the tagged two-photon production will be complementary and less model-dependant with respect to the nominal analyses of the complex final states and decay chains. In addition, the photon energy measurement will improve the reconstruction of event kinematics.

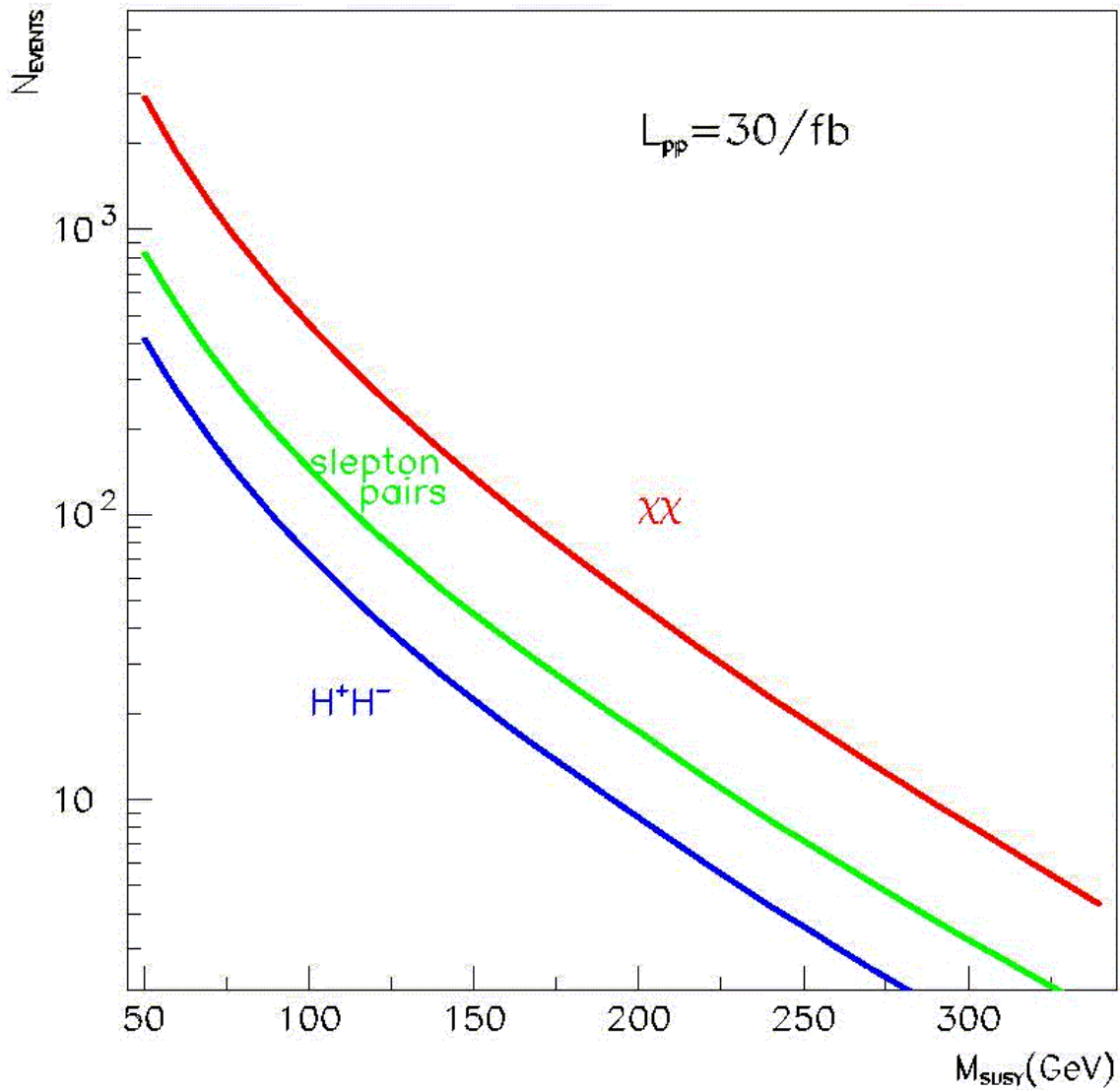

**Figure 6**. Number of pairs of the charginos, sleptons and charged Higgs bosons produced in the single tagged two-photon interactions, plotted against the mass of the produced sparticle.

In conclusion, a few new aspects of tagging two-photon production at the LHC discussed in this paper provide a further strong motivation for work towards a technical realization of this proposal. The initial survey of topics in high-energy two-photon physics shows that this research program will be a significant and complementary extension of the nominal physics program at the LHC.

Finally, the same tagging technique can also be utilized to select $\gamma p$, or $\gamma q$ and $\gamma g$, interactions at the LHC, for which the energy reach and the effective luminosity are even higher than for the $\gamma\gamma$ case. The initial study of this experimental opportunity is a subject of the publication in preparation.

## Acknowledgements

The author would like to thank the workshop organizers for support and invitation to this very interesting and inspiring conference.